\documentclass{vgtc}                          

\graphicspath{{figures/}{pictures/}{images/}{./}} 

\usepackage{times}                     

\usepackage{tabu}                      
\usepackage{booktabs}                  
\usepackage{lipsum}                    
\usepackage{mwe}                       

\usepackage{mathptmx}                  
\usepackage{tabularray}
\usepackage{siunitx}
\usepackage{gensymb}
\usepackage{color}

\onlineid{0}

\vgtccategory{Research}

\vgtcinsertpkg

\usepackage{orcidlink}

\title{Enhancement of Co-located Shared VR Experiences: Representing Non-HMD Observers on Both HMD and 2D Screens}

\author{Zixuan Guo \orcidlink{0000-0002-0451-8988}\\ %
       \parbox{1.4in}{\scriptsize \centering Xi'an Jiaotong-Liverpool University \\ The University of Liverpool} %
\and Wenge Xu \orcidlink{0000-0001-7227-7437}\\ %
    \parbox{1.4in}{\scriptsize \centering Birmingham City University} %
\and Hongyu Wang \orcidlink{0000-0002-2288-5116}\\ %
    \parbox{1.4in}{\scriptsize \centering Xi'an Jiaotong-Liverpool University} %
\and Tingjie Wan \orcidlink{0009-0003-0237-9587}\\ %
    \parbox{1.4in}{\scriptsize \centering Xi'an Jiaotong-Liverpool University \\ The University of Liverpool} %
\and Nilufar Baghaei \orcidlink{0000-0003-1776-7075}\\ %
    \parbox{1.4in}{\scriptsize \centering The University of Queensland} %
\and Cheng-Hung Lo \orcidlink{0000-0002-7199-9339}\\ %
    \parbox{1.4in}{\scriptsize \centering Xi'an Jiaotong-Liverpool University} %
\and Hai-Ning Liang~\orcidlink{0000-0003-3600-8955}\thanks{Corresponding author (e-mail: hainingliang@hkust-gz.edu.cn)}\\ %
    \parbox{1.4in}{\scriptsize \centering The Hong Kong University of Science and Technology (Guangzhou)}}

\author{
Zixuan Guo\textsuperscript{1}\textsuperscript{2}~\orcidlink{0000-0002-0451-8988}
\and 
Wenge Xu\textsuperscript{3}~\orcidlink{0000-0001-7227-7437}
\and 
Hongyu Wang\textsuperscript{1}~\orcidlink{0000-0002-2288-5116}
\and 
Tingjie Wan\textsuperscript{1}\textsuperscript{2}~\orcidlink{0009-0003-0237-9587}
\and 
Nilufar Baghaei\textsuperscript{4}~\orcidlink{0000-0003-1776-7075}
\and 
Cheng-Hung Lo\textsuperscript{5}~\orcidlink{0000-0002-7199-9339}
\and 
Hai-Ning Liang\textsuperscript{6}~\orcidlink{0000-0003-3600-8955}\thanks{Corresponding author (e-mail: hainingliang@hkust-gz.edu.cn)}
}

\affiliation{\scriptsize \textsuperscript{1} School of Advanced Technology, Xi'an Jiaotong-Liverpool University, Suzhou, China\\
\textsuperscript{2} Department of Computer Science, The University of Liverpool, Liverpool, UK\\
\textsuperscript{3} College of Computing, Birmingham City University, Birmingham, UK\\
\textsuperscript{4} School of Electrical Engineering and Computer Science, The University of Queensland, Brisbane, Australia\\
\textsuperscript{5} Department of Industrial Design, Xi'an Jiaotong-Liverpool University, Suzhou, China\\
\textsuperscript{6} Computational Media and Arts Thrust, The Hong Kong University of Science and Technology (Guangzhou), Guangzhou, China}

\teaser{
  \centering
  \includegraphics[width=\linewidth]{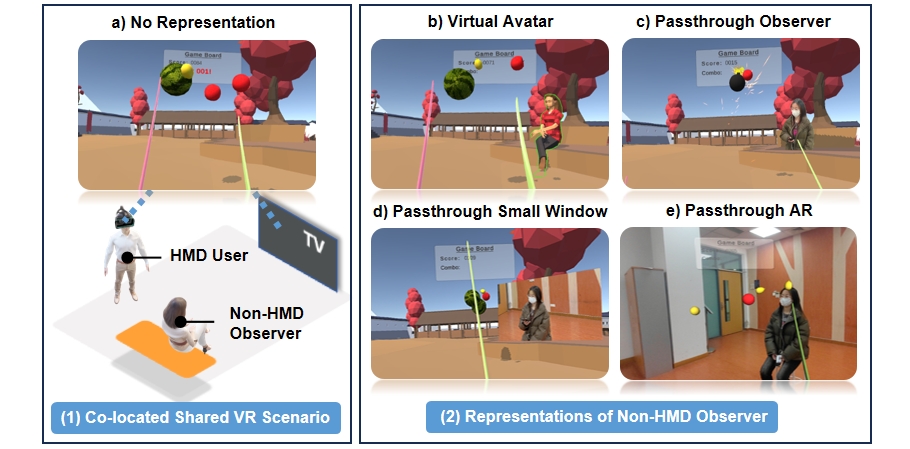}
  \caption{%
  (1) A common scenario in co-located shared VR, where the HMD user engages in VR activities, and the non-HMD observer watches these on a TV. Typically, they only see the virtual world without the non-HMD observer, as shown in (a) No Representation (Baseline). (2) The 4 methods used in our study to represent the non-HMD observer in the virtual world: (b) Virtual Avatar, (c) Passthrough Observer, (d) Passthrough Small Window, and (e) Passthrough Augmented Reality. %
  }
  \label{fig:teaser}
}

\abstract{%
Virtual reality (VR) not only allows head-mounted display (HMD) users to immerse themselves in virtual worlds but also to share them with others. When designed correctly, this shared experience can be enjoyable. However, in typical scenarios, HMD users are isolated by their devices, and non-HMD observers lack connection with the virtual world. To address this, our research investigates visually representing observers on both HMD and 2D screens to enhance shared experiences. The study, including five representation conditions, reveals that incorporating observer representation positively impacts both HMD users and observers. For how to design and represent them, our work shows that HMD users prefer methods displaying real-world visuals, while observers exhibit diverse preferences regarding being represented with real or virtual images. We provide design guidelines tailored to both displays, offering valuable insights to enhance co-located shared VR experiences for HMD users and non-HMD observers.
}


\begin{document}



\firstsection{Introduction}

\maketitle

Virtual reality (VR) technologies offer highly immersive environments through head-mounted displays (HMDs). However, HMD users often struggle to balance immersion in the virtual world with interacting with others in the same shared space. For instance, during shared gaming experiences, HMD users explore virtual worlds while observers watch and engage in the excitement and strategy of the game \cite{krekhov2020silhouette, thoravi2022dreamstream}. This scenario also applies to technical demonstrations, where HMD users showcase various VR applications and artworks to observers, highlighting the creativity and practical uses of VR \cite{welsford2021spectator, thoravi2022dreamstream, kamei2020covr}. In educational or professional settings, VR simulates real-world experiences and training scenarios, allowing HMD users to educate and train observers effectively through interactive VR experiences \cite{thoravi2022dreamstream, kamei2020covr}. In these common scenarios, HMD users are immersed in the virtual worlds, while those without HMDs assume the role of observers, watching the VR content that is projected through traditional 2D screens like TVs and computer monitors. 

However, in these shared contexts, both HMD users and non-HMD observers face distinct challenges. The enclosed nature of HMDs poses significant obstacles for HMD users in perceiving observers and interacting with them \cite{ghosh2018notifivr, o2023re}. Furthermore, HMD users' limited visibility of the real world in the presence of others raises issues related to personal privacy and social comfort \cite{o2021safety, osmers2021role, rettinger2022you, dao2021bad}. For non-HMD observers, as they can only watch through 2D screens, they may experience a diminished sense of connection with the virtual world, resulting in lower overall enjoyment in this shared experience \cite{welsford2021spectator, gugenheimer2017sharevr}.

Some studies have attempted to address the challenges faced by HMD users and observers. For HMD users, research \cite{ghosh2018notifivr, o2020reality, rzayev2019notification, o2022exploring, george2018intelligent, ghosh2018notifivr, gottsacker2021diegetic, medeiros2021promoting, kudo2021towards, george2020seamless, o2023re} has proposed various reality awareness systems to represent observers in VR, with findings indicating that HMD users prefer more visual cues during prolonged interactions to enhance their awareness of the actual environment. However, previous studies \cite{o2023re, gottsacker2021diegetic, kudo2021towards} have predominantly used these representations as cues for when others approach or interrupt. For non-HMD observers, there is currently no systematic solution for enhancing their watching experience. Several studies \cite{gugenheimer2017sharevr, welsford2021spectator, lee2020rolevr, emerson2021enabling, kamei2020covr} aimed to elevate observer engagement by introducing interactive features; however, these innovative methods require additional development efforts and come with higher costs and complex setups, presenting significant challenges for practical implementation in real-world scenarios.

In response to these gaps, our study aims to explore how non-HMD observers can be visually represented on both HMD and 2D screens to enhance the shared experience between HMD users and non-HMD observers. As shown in Figure \ref{fig:teaser}, we conducted an experiment with five conditions: No Representation (Baseline), Virtual Avatar (VA), Passthrough Observer (PO), Passthrough Small Window (PSW), and Passthrough Augmented Reality (PAR). 40 participants, organized into pairs of HMD users and observers, participated in the experiment. We utilized a common real-life scenario involving prolonged interaction periods in shared spaces, specifically game-sharing in a living room \cite{gugenheimer2017sharevr, venkatesh2003networked, xu2024exploring}. The study shows that HMD users prefer real-world visuals of observers, with PO being their top choice. Non-HMD observers favor PO and VA for their realistic and virtual representations, respectively. Our study provides design guidelines for observer representation, emphasizing tailored approaches for different perspectives. 

Overall, our paper makes the following contributions:

\begin{itemize}
\item A first exploration of the influence of visually representing observers in shared VR scenarios.
\item A systematic evaluation from the perspectives of both HMD users and observers.
\item A set of design recommendations for representing observers on VR HMDs and 2D screens.
\end{itemize}

\section{Related Work}
\subsection{Challenges of Co-Located Shared VR}
While it is common for HMD users to share VR content with co-located non-HMD observers, each party faces different challenges. For HMD users, the immersive nature of HMDs creates a substantial barrier, hindering their awareness of observers and impeding effective communication \cite{o2023re, osmers2021role, maurer2015gaze, o2021safety}. The obscuring or loss of social cues, crucial for human interaction \cite{polzin2000verbal, argyle1970communication}, in HMD usage may result in miscommunication and a lack of nuanced social interaction, potentially inducing social isolation \cite{liszio2017influence, jansen2020share, gugenheimer2017sharevr, kamei2020covr, matsuda2021vr}. Furthermore, privacy and comfort issues may arise, with HMD users in a vulnerable position, raising apprehensions about unexpected behavior from observers, such as filming and pranks \cite{o2021safety, osmers2021role, rettinger2022you, dao2021bad}.

For observers watching through conventional 2D screens, they also face a set of challenges. Primarily in a passive role, observers may experience a diminished sense of connection to the virtual world, leading to potential boredom and reduced enjoyment \cite{welsford2021spectator, lee2020rolevr, sra2016resolving, gugenheimer2017sharevr}. Additionally, observers may find it challenging to effectively convey their feelings and thoughts to HMD users, significantly impacting their shared experience \cite{kamei2020covr, dao2021bad, mai2017transparenthmd}.

In summary, both HMD users and observers face challenges in shared spaces: HMD users need to overcome the isolation imposed by HMDs, while observers require an enhanced connection to the virtual world. Incorporating interactive technologies on both HMD and traditional 2D screens is crucial for addressing these issues and enhancing their shared experience.

\subsection{Representing Non-HMD Observers to Enhance HMD Users' Awareness}
Addressing concerns about isolation caused by HMDs, research has focused on improving user awareness by representing non-HMD observers in the virtual environment. Various methods have been proposed, including text notifications \cite{ghosh2018notifivr, o2020reality, rzayev2019notification}, audio notifications \cite{o2020reality, o2022exploring}, haptic notifications \cite{george2018intelligent, ghosh2018notifivr}, avatar designs \cite{gottsacker2021diegetic, medeiros2021promoting, kudo2021towards, george2020seamless}, symbol designs \cite{medeiros2021promoting, kudo2021towards}, and Passthrough functions \cite{gottsacker2021diegetic, von2019you, mcgill2015dose, george2020seamless, guo2024exploring}. O'Hagan et al. \cite{o2023re} assessed awareness systems and interaction scenarios with observers, highlighting a preference for visual representations and the need for more visual cues during prolonged communication.

Among the visual solutions, Passthrough-based techniques can provide HMD users with an authentic view of their surroundings. These methods encompass varying degrees of real-world visuals, commonly including Passthrough Observer (PO), which only displays the observer \cite{o2023re, von2019you, mcgill2015dose}; Passthrough Small Window (PSW) that presents the real scene in a windowed format \cite{wang2022realitylens, von2019you}; and Passthrough Augmented Reality (PAR) \cite{o2023re, george2020seamless, guo2024breaking} showing an AR version of the application.

Virtual representations, mainly through avatars and symbols, allow natural integration with the virtual environment. Research suggests that avatars are more effective than symbols in representing observers \cite{kudo2021towards, medeiros2021promoting}. Gottsacker et al. \cite{gottsacker2021diegetic} emphasized that avatars with a thematically similar appearance to the virtual environment enhance the interactive experience with minimal disruption.

While effective for brief cues, these methods have seen limited exploration for prolonged interaction in shared spaces. O'Hagan et al. \cite{o2022exploring} delved into audio methods for facilitating communication between HMD users and non-HMD observers, but visual methods remain underexplored. Our research aims to explore real-time visual representations of non-HMD observers in VR during prolonged interactions.

\subsection{Enhancing Non-HMD Observers' Connection to the Virtual World}
Enhancing the non-HMD observer's watching experience crucially depends on strengthening their connection with the virtual world; however, there is currently a lack of a systematic solution. Most research efforts \cite{gugenheimer2017sharevr, welsford2021spectator, emerson2021enabling, kamei2020covr, lee2020rolevr} focus on increasing observer interaction with the virtual world. For instance, Gugenheimer et al. \cite{gugenheimer2017sharevr} proposed ShareVR, allowing observers to watch the virtual world on a floor projection and actively participate in the gaming experience. Emerson et al. \cite{emerson2021enabling} enabled observers to watch VR content on a large display wall and annotate using text.

Some studies \cite{krekhov2020silhouette, ishii2019let, emmerich2021streaming} focus on displaying HMD users in the virtual world from a third-person perspective, allowing observers to see their real bodies and movements. For example, Krekhov et al. \cite{krekhov2020silhouette} created a one-way mirror, enabling observers to see the HMD user's reflection in the virtual world. Ishii et al. \cite{ishii2019let} introduced ReverseCAVE, allowing observers to see both the HMD user and the VR environment on semi-transparent screens.

While many methods increase development complexity and costs, third-person perspectives are more feasible. However, Emmerich et al. \cite{emmerich2021streaming} suggest observers prefer a first-person perspective for better focus and immersion. Consequently, in this study, we use first-person perspective to explore the impact of representing observers themselves in the virtual world—a novel approach that, to our knowledge, has not been previously investigated.

\section{Experiment}
\subsection{Experiment Design}
Our experiment involved two participants: one as the HMD user and the other as the non-HMD observer. We chose to simulate a scenario of game sharing in a living room, which represents a common shared space activity in daily life where HMD users and observers interact \cite{gugenheimer2017sharevr, jones2014roomalive, venkatesh2003networked}. In this setup, the HMD user played the game while the observer watched on the television screen. According to \cite{emmerich2017impact}, shared gaming requires players to focus visually on the game while interacting audibly with others, challenging their attention and working memory. Thus, we included the N-back task as a secondary task for HMD users.

Based on a review of the literature \cite{gottsacker2021diegetic, von2019you, o2023re, mcgill2015dose, kudo2021towards, medeiros2021promoting, wang2022realitylens}, we implemented four visual representations of non-HMD observers with varying degrees of physical realism, plus a No Representation (Baseline) condition, creating five experimental conditions (Figure \ref{fig:teaser}). In each condition, the visuals on the TV screen mirrored those within the VR environment, providing a first-person perspective, allowing both the HMD user and the non-HMD observer to experience all five conditions collectively, with a counterbalanced sequence. Pilot testing with four testers ensured consistent observer size and positioning across conditions, and no latency issues were reported. We developed the program for this experiment using Unity3D, version 2021.3.26f1. The specifics of each condition are detailed below:

\begin{itemize}
\item \textbf{No Representation (Baseline):} In this baseline condition, HMD users can only see the game visuals without any representation of non-HMD observers.

\item \textbf{Virtual Avatar (VA):} According to \cite{gottsacker2021diegetic}, we employed a virtual avatar featuring a low-poly style and integrated a green halo effect.
Participants have the option to choose between a male and a female avatar. To enhance the natural and lively appearance of the avatar, we implemented three animations--the avatar sitting still, sitting but gesturing with hands, and a third sitting with cross-legged. These three animations play in a random sequence, one after another.

\item \textbf{Passthrough Observer (PO):} Building upon \cite{o2023re, mcgill2015dose, von2019you}, this method presents a passthrough view that removes the background, displaying only the observers. We first captured real-time video of the observer using a camera in front of the HMD. Then, we used portrait segmentation via Tencent meeting to isolate the observer from the background. Finally, we integrated the processed video into the virtual environment. 

\item \textbf{Passthrough Small Window (PSW):} As outlined in \cite{wang2022realitylens, von2019you}, this approach displays the Passthrough video in a window format, offering a view of both the observer and their surroundings. Mirroring the implementation of PO, we first captured real-time video of the observer through an additional camera. This video, without additional processing, was placed in the virtual environment with the same position as in the PO condition.

\item \textbf{Passthrough Augmented Reality (PAR):} This method includes switching to an AR version of the application, preserving only essential in-game objects while replacing the rest with the Passthrough view \cite{o2023re, mcgill2015dose}.

\end{itemize}

\subsection{Apparatus and Setup}
We utilized a Pico 4 as our VR HMD for the HMD user, along with a 50-inch 4K MI TV, model L50M5-AD, as the display for observers. To capture real-world scenes integrated into the virtual environment, we employed the AUSDOM AW651 Webcam, a 1080P 60FPS camera fixed to the front of the VR HMD \cite{mcgill2015dose}. All these devices were linked to a Windows 10 PC with an i7-8700K CPU and a GTX 1080 GPU. Additionally, the observer was provided with a Google Pixel 3 connected to headphones to receive the audio information for the N-back task.

\begin{figure}[htbp]
    \centering
    \includegraphics[width=\columnwidth]{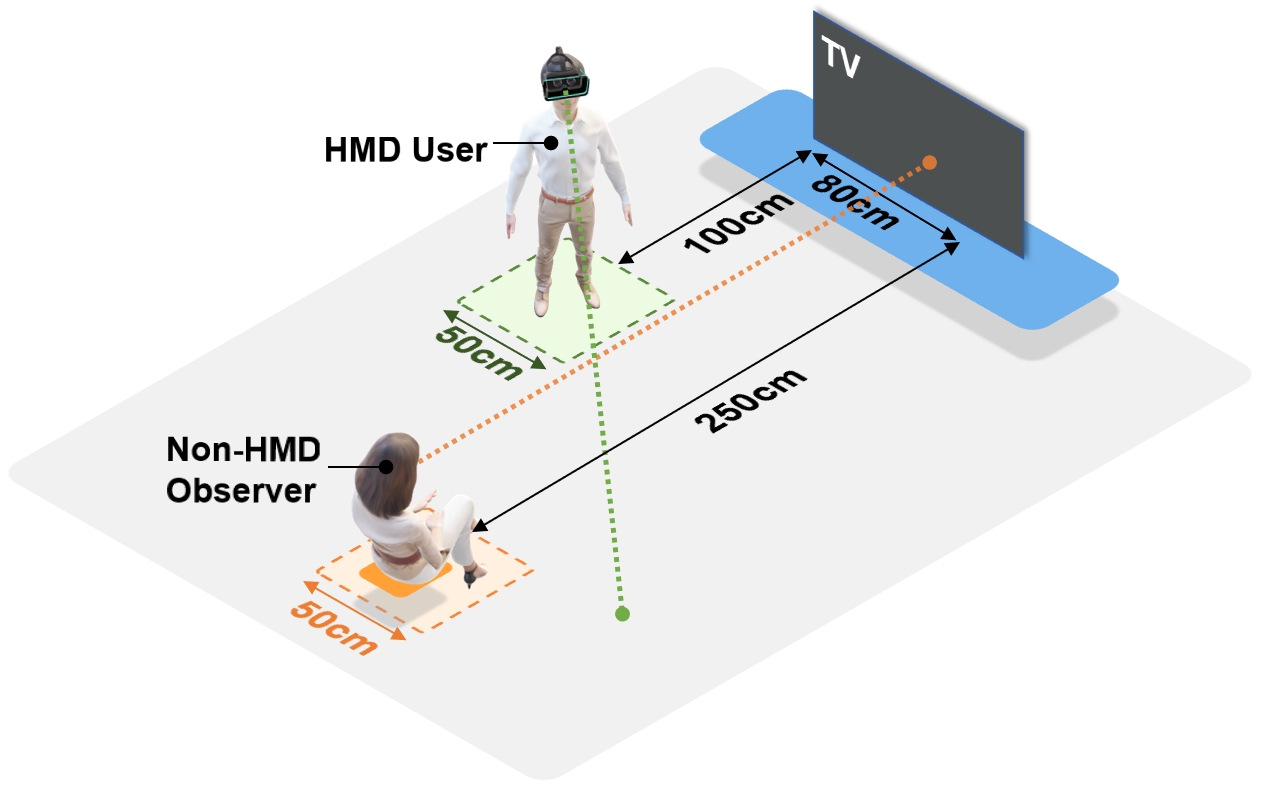}
    \caption{Living room-like environment with the TV, HMD user, and non-HMD observer positions during the experiment.}
    \label{Setup}
\end{figure}

As shown in Figure \ref{Setup}, we recreated a living room-like setting. During the experiment, the observer occupied a fixed seat approximately \SI{250}{cm} away from the TV, facing the screen. The HMD user, on the other hand, was situated about \SI{100}{cm} away from the TV, facing a direction approximately 30\degree \;away from the observer's orientation, resulting in the observer appearing in the lower-right corner of the HMD user's field of view. The horizontal distance between the observer and the HMD user was approximately \SI{80}{cm}. The fixed areas for both the observer and the HMD user were around \SI{50}{cm^2}, accommodating participants of varying heights and providing the necessary space for gameplay.

This experiment was conducted in a controlled laboratory environment, ensuring adequate lighting and no external disruptions. Close supervision by an experimenter was maintained throughout to mitigate potential risks. This study received ethical approval from the University Ethics Committee at our institution.

\subsection{Primary Task}
We used a game as the primary task within the virtual environment for the experiment. Firstly, games are representative of typical VR consumer applications, aligning with the prevalent use of gaming scenarios in past research on awareness systems \cite{o2023re, von2019you}. Secondly, games foster a shared experience for both HMD users and non-HMD observers, promoting interaction and enjoyment. 

We developed a game similar to VR Fruit Ninja \footnote{https://store.steampowered.com/app/486780/Fruit\_Ninja\_VR/} and used a low-poly Japanese-style courtyard as the game scene. Players utilize two handheld controllers to wield virtual swords in the game, aiming to slice as many fruits (watermelon, apple, and lemon) as possible while avoiding incoming bombs. Building on the game used in \cite{guo2023s, guo2024breaking}, fruits and bombs are launched from both the left and right sides of the player, following a parabolic trajectory. The game spans 5 minutes, divided into five 1-minute sequences. Each sequence comprises 30 rapid 2-second rounds, featuring a dynamic mix of fruits and bombs. Each round generates 2-4 objects, with a bomb appearing every 5 rounds. 



\subsection{Secondary Task}
Sharing the VR experience with observers means that HMD users need to complete not only the primary tasks in the virtual world but also interact with the observers. In this experiment, we use a variant of the N-back task, commonly employed in driving research, as the secondary task to assess working memory and attentional shifts \cite{mehler2012sensitivity, he2019high, mehler2009impact}. 
In this task, participants listen to a sequence of 10 non-repeating single-digit numbers (0–9) presented randomly as auditory stimuli. After each number, they immediately recall and verbally respond with the number presented n positions back in the sequence. For example, in the 1-back task, participants respond with the number one position back, while in the 2-back task, they recall the number two positions back.

Given this scenario, where HMD users need to interact with observers, we required the observers to present the HMD user with the cues for the N-back task to better simulate this interaction. Observers listened to pre-recorded audio through headphones and repeated the difficulty cue and numbers out loud to HMD users immediately after hearing them. Each condition included four N-back tasks (two 1-back and two 2-back tasks) with a balanced task order. Following prior research \cite{mehler2012sensitivity, mehler2009impact}, each N-back task comprised 10 numbers (0-9) spaced 2.25 seconds apart. A difficulty cue, followed by a 2.25-second interval, signaled the start of each task, lasting about 22.5 seconds. There was a 30-second interval between tasks, and the first task began one minute after the game started.

\subsection{Procedure}
Upon arrival, participants were briefed on the experiment's purpose and procedures and provided with consent forms to review and sign. They were randomly assigned roles—one as the HMD user and the other as the non-HMD observer. Both completed a pre-experiment questionnaire.

The N-back task was introduced, and participants completed practice sessions with 1-back and 2-back tasks to ensure comprehension. The HMD user received gameplay instructions and a one-minute training session without representation techniques to familiarize themselves with the equipment and gameplay.

As shown in Figure \ref{Setup}, the experimenter positioned the observer and HMD user to ensure clear visibility. Prior to each condition, the experimenter checked and adjusted participants' positions, assisted with the VR HMD and headphones, and ensured no communication occurred between participants during the experiment, except for the N-back task. Accuracy of N-back responses was recorded manually, and the procedure was video-recorded for later review.

After each condition, both participants completed post-experiment questionnaires. At the end of the experiment, a joint semi-structured interview was conducted to rank the conditions and gather feedback. Each participant's total involvement in the experiment was approximately 60 minutes.

\subsection{Participants}
We recruited participants in pairs, all of whom were already familiar with each other. A total of 20 pairs, comprising 40 participants (16 females and 24 males; mean age = 23.2, SD = 2.3), were recruited. Among the 20 HMD users, 14 had previous experience using VR HMDs, with 5 of them being regular weekly users. Among the 20 observers, 16 had prior experience using VR HMDs. Regarding watching others use VR, 3 did so frequently, and 6 had some experience. All participants volunteered for the study without compensation.

\subsection{Measures}
We conducted measurements for both HMD users and non-HMD observers, with only Task Performance and Cybersickness requiring separate evaluation for HMD users.
\subsubsection{Measures for HMD Users}

\begin{itemize}
\item \emph{\textbf{Task Performance.}} We collected data on the HMD user's gaming performance, including: (1) game score; (2) success rates in slicing fruits and avoiding bombs; and (3) maximum combo count. For communication performance, we also recorded their (4) accuracy rates in the 1-back and 2-back tasks.

\item \emph{\textbf{Cybersickness.}} The level of cybersickness experienced by HMD users was measured by the Simulator Sickness Questionnaire (SSQ) \cite{kennedy1993simulator}, which consists of 16 items graded on a scale from 0 (none) to 3 (severe). It assesses nausea, oculomotor discomfort, and disorientation. 

\end{itemize}

\subsubsection{Measures for Both HMD Users and Non-HMD Observers}

\begin{itemize}

\item \emph{\textbf{System Usability.}} We followed \cite{o2023re} to assess the system usability by asking both HMD users and observers to what extent they agreed on these 7 statements: (1) ``was disruptive”, (2) ``was frustrating”, (3) ``felt natural”, (4) ``was easy to understand”, (5) ``was informative”, (6) ``improved your ability to communicate”, and (7) ``made you too aware of the real world” (not applicable in the observer version). These statements were rated in a 7-point Likert scale.

\item \emph{\textbf{Presence Experience.}} We measured HMD users and observers on their sense of presence in the virtual environment using the ``Sense of Being There'' subscale of the Igroup Presence Questionnaire (IPQ) \cite{schubert2001experience}, consisting of a single item. HMD users were additionally assessed for their attention level and awareness of observers' presence using the 4-item ``Involvement'' subscale of IPQ \cite{schubert2001experience} and 3 items from the ``Co-presence'' subscale of the Networked Minds Social Presence Measure \cite{harms2004internal}. All these scales were measured on a 7-point Likert scale.

\item \emph{\textbf{Emotions.}} For both HMD users and observers, we employed the valence and arousal dimensions from the Self-Assessment Manikin (SAM) scale \cite{bradley1994measuring}, gauging valence (from unpleasant to pleasant) and arousal (from calm to excited) using a 7-point scale.

\item \emph{\textbf{Semi-structured Interview.}} Both HMD users and observers were asked to rank five conditions and provide reasons for their choices. Interviews were audio-recorded and transcribed for data analysis.

\end{itemize}

\subsection{Hypotheses}
For the perspective of HMD users, previous studies \cite{o2023re, gottsacker2021diegetic, kudo2021towards, von2019you} focusing on cues for HMD users suggest that representing observers in the virtual environment can enhance HMD users' awareness of the observers and increase the sense of co-presence. According to the systematic evaluation by O'Hagan et al. \cite{o2023re}, representations involving Passthrough functionality can provide more real-world information and facilitate communication, but they may reduce the sense of immersion. From the observers' perspective, previous research \cite{gugenheimer2017sharevr, lee2020rolevr} suggests that their sense of presence and overall enjoyment increase when they can engage with virtual activities through various technologies. However, the ability to see both real and virtual content simultaneously may lead to distractions \cite{jansen2020share}. Based on these findings from studies focusing on HMD users, we propose the following hypotheses:

\begin{itemize}

\item {\textbf{H1.}} The presence of representations, compared to their absence, will {\textbf{(a)}} enhance HMD user's attention to the interaction with observer and {\textbf{(b)}} increase the sense of co-presence with the observer, {\textbf{(c)}} providing a more enjoyable shared experience.

\item {\textbf{H2.}} The representations with Passthrough, compared to without, enable HMD user to {\textbf{(a)}} perceive more information, {\textbf{(b)}} facilitate communication, but {\textbf{(c)}} diminish their sense of presence in the virtual world.

\item {\textbf{H3.}} The presence of representations, compared to their absence, will {\textbf{(a)}} enhance the observer's sense of presence in the virtual world and {\textbf{(b)}} provide a more enjoyable shared experience.

\item {\textbf{H4.}} The representations with Passthrough, compared to without, allow the observer to {\textbf{(a)}} perceive more information, but {\textbf{(b)}} introduce more distractions.

\end{itemize}

\section{Results}
Prior to analysis, we examined the normal distribution of the data using Shapiro-Wilk tests and Q-Q plots. For non-normally distributed data, we utilized the Aligned Rank Transform (ART) \cite{wobbrock2011aligned} to perform transformations before conducting ANOVAs. We conducted one-way repeated-measures ANOVAs (RM-ANOVA) and applied Bonferroni corrections for all pairwise comparisons, reporting effect sizes wherever feasible. When Mauchly’s test indicated violation of the assumption of sphericity, degrees of freedom were adjusted using Greenhouse-Geisser estimates. 

\subsection{Task Performance}
\begin{figure*}[htbp]
    \centering
    \includegraphics[width=\textwidth]{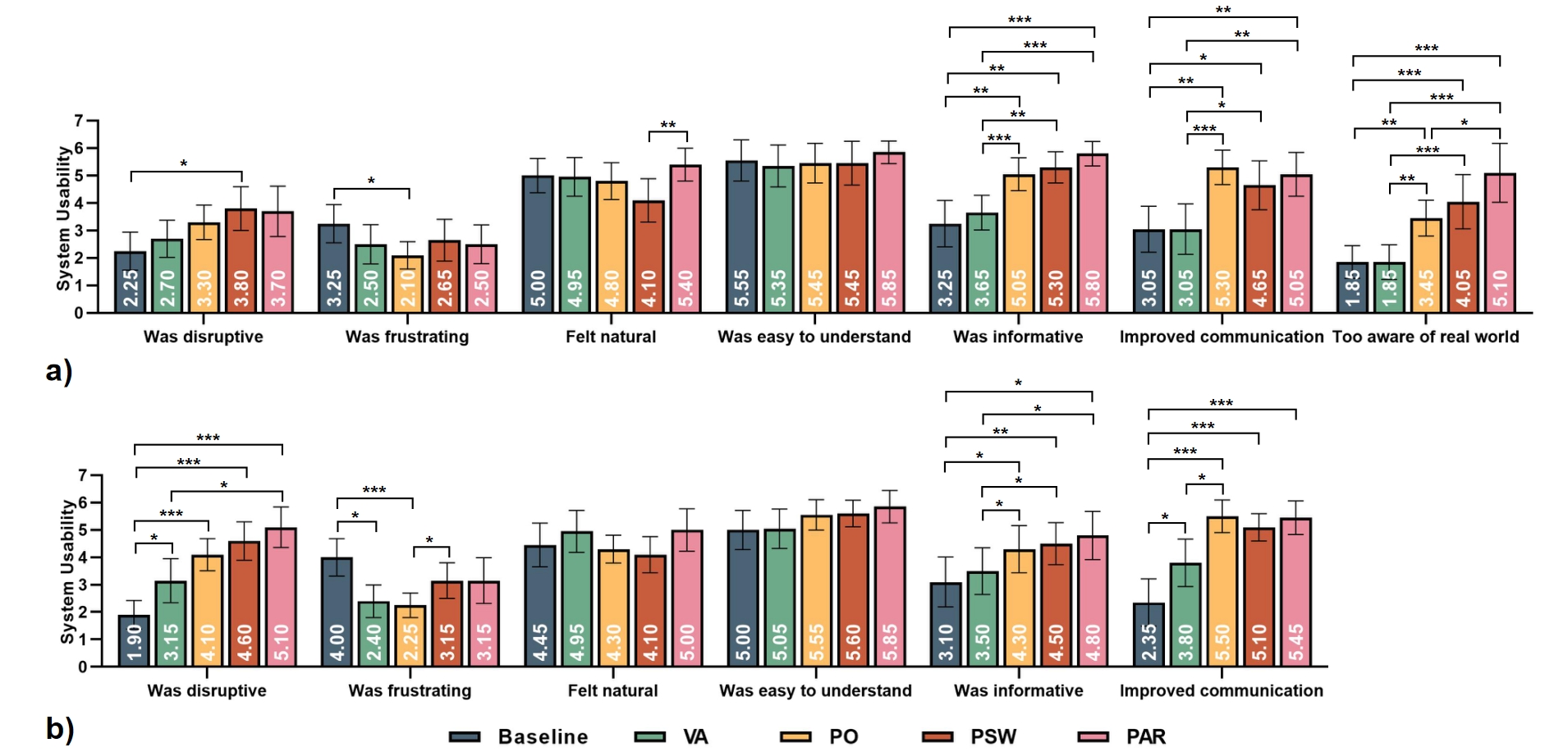}
    \caption{System usability ratings for (a) HMD users and (b) non-HMD observers in each condition. Error bars indicate 95\% confidence intervals. *, **, and *** indicate statistical significance at the $p < 0.05, p < 0.01,$ and $p < 0.001$ levels, respectively.}
    \label{Usability}
\end{figure*}

For N-back tasks, we found significant differences in the accuracy rates in the 2-back tasks ($F_{2.647, 50.297}=3.086, p=.041, \eta_p^2=0.140$). Post-hoc tests indicated a significantly higher accuracy for HMD users in the PO condition ($M=98.44\%, SD=0.03$) compared to the Baseline ($M=92.81\%, SD=0.08) (p=.007$) and VA ($M=90.94\%, SD=0.11) (p=.004$). We did not observe significant differences in the accuracy rates among the conditions in the 1-back tasks, nor in any measures of game performance.

\subsection{Cybersickness}
None of the HMD users in our study exhibited SSQ scores exceeding 30 across the five conditions: Baseline ($M=1.87, SD=2.57$), VA ($M=2.06, SD=2.84$), PO ($M=2.24, SD=2.55$), PSW ($M=2.06, SD=2.57$), and PAR ($M=2.24, SD=2.82$). Furthermore, the results did not reveal any significant differences in scores related to Nausea, Oculomotor, Disorientation, or the overall total SSQ scores across the five conditions.

\subsection{System Usability}

\subsubsection{HMD users}
The results of usability evaluation are summarised in Figure \ref{Usability}. A significant effect of condition on disruptive was observed ($F_{4,76}=5.687, p<.001, \eta_p^2=0.230$). The post-hoc tests indicated that PSW ($M=3.80, SD=1.70$) led to significantly greater disruption compared to Baseline ($M=2.25, SD=1.48) (p=.017$).

In terms of frustration, a significant effect of condition was identified ($F_{4,76}=3.067, p=.021, \eta_p^2=0.139$). Post-hoc tests exposed a significant increase in frustration within the Baseline condition ($M=3.25, SD=1.48$) compared to PO ($M=2.10, SD=1.07) (p=.019$).

Regarding the perception of naturalness, a significant effect of condition was observed ($F_{4,76}=3.327, p=.014, \eta_p^2=0.149$). Post-hoc tests indicated that PAR ($M=5.40, SD=1.27$) is a significantly more natural technique compared to PSW ($M=4.10, SD=1.68) (p=.001$).

For informativeness, a significant effect of condition was observed ($F_{4,76}=19.134, p<.001, \eta_p^2=0.502$). Subsequent post-hoc tests revealed that Baseline ($M=3.25, SD=0.40$) provided less information compared to PO ($M=5.05, SD=1.28) (p=.007$), PSW ($M=5.30, SD=1.22) (p=.001$), and PAR ($M=5.80, SD=0.95) (p<.001$). Additionally, VA ($M=3.65, SD=1.35$) also offered less information than PO ($p<.001$), PSW ($p=.002$), and PAR ($p<.001$).

Concerning improving communication, a significant effect of condition was observed ($F_{4,76}=13.191, p<.001, \eta_p^2=0.410$). Similarly to above, Baseline ($M=3.05, SD=1.79$) yielded worse communication enhancement compared to PO ($M=5.30, SD=1.34) (p=.001$), PSW ($M=4.65, SD=1.90) (p=.043$), and PAR ($M=5.05, SD=1.70) (p=.001$). VA ($M=3.05, SD=1.96$) showed worse communication improvement than PO ($p<.001$), PSW ($p=.021$), and PAR ($p=.002$).

Furthermore, we observed a significant effect of condition regarding the perception of being too aware of the real world ($F_{4,76}=22.756, p<.001, \eta_p^2=0.545$). Our post-hoc tests indicated that, compared to Baseline ($M=1.85, SD=1.27$), HMD users were more aware of the real world in PO ($M=3.45, SD=1.40) (p=.001$), PSW ($M=4.05, SD=2.11) (p<.001$), and PAR ($M=5.10, SD=2.30) (p<.001$). In contrast to VA ($M=1.85, SD=1.35$), HMD users also demonstrated heightened awareness in PO ($p=.001$), PSW ($p<.001$), and PAR ($p<.001$). Furthermore, HMD users exhibited increased awareness in PAR compared to PO ($p=.046$). No significant differences in the perception of being easy to understand were observed among the conditions.

\subsubsection{Non-HMD Observers}
As shown in Figure \ref{Usability}, we observed a significant effect of condition on disruption ($F_{2.424,46.064}=17.880, p<.001, \eta_p^2=0.485$). Post-hoc tests indicated that the Baseline ($M=1.90, SD=1.12$) caused significantly less disruption to observers compared to the VA ($M=3.15, SD=1.73) (p=.023$), PO ($M=4.10, SD=1.25, p<.001$), PSW ($M=4.60, SD=1.50, p<.001$), and PAR ($M=5.10, SD=1.59, p<.001$). Additionally, VA resulted in significantly less disruption than PAR ($p=.016$).

For frustration, we identified a significant effect of condition ($F_{2.574,48.901}=7.896, p<.001, \eta_p^2=0.294$). The post-hoc tests revealed that the Baseline ($M=4.00, SD=1.45$) induced significantly more frustration than both VA ($M=2.40, SD=1.27) (p=.023$) and PO ($M=2.25, SD=0.97, p<.001$). Additionally, PSW ($M=3.15, SD=1.39$) also led to more frustration among observers compared to PO ($p=.019$).

Concerning informativeness, we found a significant effect of condition ($F_{2.538,48.217}=9.538, p<.001, \eta_p^2=0.334$). Post-hoc tests revealed that Baseline ($M=3.10, SD=1.94$) provided less information compared to PO ($M=4.30, SD=1.84) (p=.025$), PSW ($M=4.50, SD=1.64) (p=.008$), and PAR ($M=4.80, SD=1.88) (p=.02$). Additionally, VA ($M=3.50, SD=1.82$) also yielded less information than PO ($p=.031$), PSW ($p=.016$), and PAR ($p=.017$).

In terms of improving communication, a significant effect of condition was found ($F_{2.417,45.916}=19.299, p<.001, \eta_p^2=0.504$). Post-hoc tests highlighted that the VA ($M=3.80, SD=1.85) (p=.03$), PO ($M=5.50, SD=1.28, p<.001$), PSW ($M=5.10, SD=1.07, p<.001$), and PAR ($M=5.45, SD=1.32, p<.001$) were more effective than the Baseline ($M=2.35, SD=1.84$) in facilitating communication. Furthermore, PO demonstrated a significantly greater ability to enhance communication compared to VA ($p=.019$). There were no significant differences observed among the conditions in terms of the perception of naturalness and the perception of being easy to understand.

\subsection{Presence Experience}

\subsubsection{HMD users}
HMD users' presence experiences in each condition are shown in Figure \ref{Presence}. A significant effect of condition on HMD users' perceived presence in virtual environment was observed ($F_{4,76}=12.479, p<.001, \eta_p^2=0.396$). Post-hoc tests showed that Baseline ($M=6.20, SD=0.87$) was significantly higher than PO ($M=4.75, SD=1.30, p<.001$), PSW ($M=4.30, SD=1.38, p<.001$), and PAR ($M=5.10, SD=0.89) (p=.001$). PSW was significantly lower than both VA ($M=5.75, SD=1.34) (p=.017$) and PAR ($p=.028$).

In terms of involvement, a significant effect was found for condition ($F_{2.327,44.205}=20.630, p<.001, \eta_p^2=0.521$). Post-hoc tests indicated that PAR  ($M=2.84, SD=0.76$) demonstrated a significant decrease compared to Baseline ($M=4.69, SD=0.79, p<.001$), VA ($M=4.58, SD=0.79, p<.001$), PO ($M=3.90, SD=0.84, p<.001$), and PSW ($M=3.56, SD=1.16) (p=.013$). Furthermore, PO ($p=.01$) and PSW ($p=.007$) exhibiting significant decreases compared to Baseline.

For their perceived co-presence, significant differences between conditions were observed ($F_{4,76}=25.475, p<.001, \eta_p^2=0.573$). Post-hoc tests revealed that Baseline ($M=2.85, SD=1.43$) showed a significant decrease compared to PO ($M=5.43, SD=1.42, p<.001$), PSW ($M=5.30, SD=1.47) (p=.001$), and PAR ($M=6.08, SD=0.10, p<.001$). Similarly, VA ($M=3.27, SD=1.69$) also exhibited a significant decrease compared to PO ($p<.001$), PSW ($p=.003$), and PAR ($p<.001$). 

\subsubsection{Non-HMD Observers}
\begin{figure*}[htbp]
    \centering
    \includegraphics[width=\textwidth]{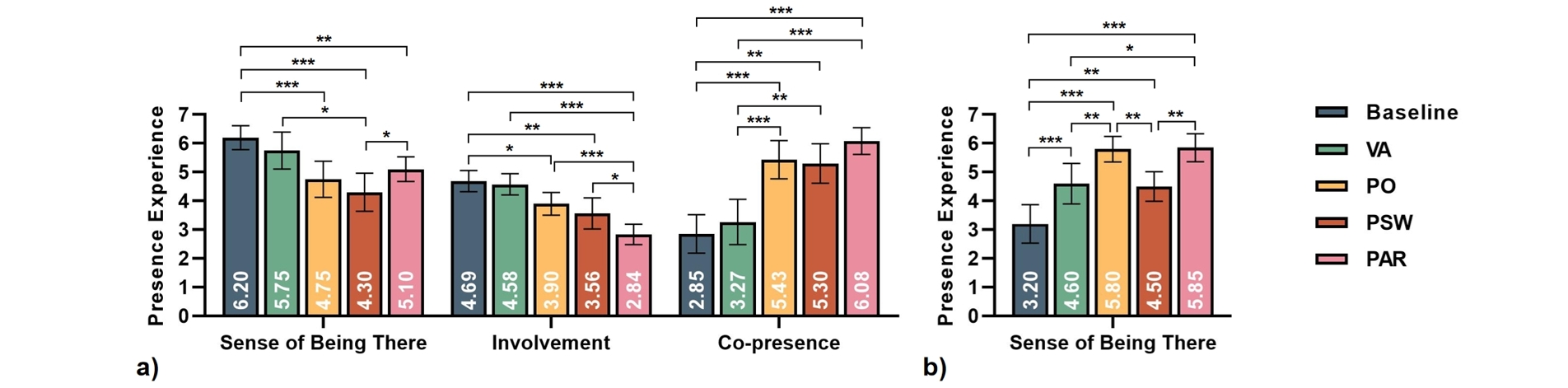}
    \caption{(a) Sense of Being There, Involvement, and Co-presence ratings for HMD users in each condition. (b) Sense of Being There ratings for non-HMD observers in each condition. Error bars indicate 95\% confidence intervals. *, **, and *** indicate statistical significance at the $p < 0.05, p < 0.01,$ and $p < 0.001$ levels, respectively.}
    \label{Presence}
\end{figure*}

As shown in Figure \ref{Presence}, we found significant differences for observers' perceived presence in the virtual environment across the five conditions ($F_{4,76}=27.033, p<.001, \eta_p^2=0.587$). Post-hoc analyses indicated that Baseline ($M=3.20, SD=1.44$) scored lower than VA ($M=4.60, SD=1.50, p<.001$), PO ($M=5.80, SD=0.95, p<.001$), PSW ($M=4.50, SD=1.10) (p=.001$), and PAR ($M=5.85, SD=1.04, p<.001$). PSW scored lower than PO ($p=.002$) and PAR ($p=.003$), and VA also lower than PO ($p=.007$) and PAR ($p=.016$).

\subsection{Emotions}

\subsubsection{HMD users}
As illustrated in Figure \ref{SAM}, results showed that HMD users’ perceived valence changed significantly across the five conditions ($F_{4,76}=9.652, p<.001, \eta_p^2=0.337$). The post-hoc tests revealed that the valence score of PO ($M=5.85, SD=1.18$) was significantly higher than that of Baseline ($M=4.65, SD=1.23) (p=.001$), VA ($M=4.80, SD=1.44) (p=.012$), and PSW ($M=5.20, SD=1.32) (p=.02$); whereas the valence score of PAR ($M=5.40, SD=1.14$) was significantly higher than Baseline ($p=.026$).

In the context of the arousal dimension, a significant effect of condition was identified ($F_{2.865,54.438}=4.441, p=.008, \eta_p^2=0.189$). Subsequent post-hoc tests revealed that the arousal score of PO ($M=5.50, SD=1.24$) significantly exceeded that of Baseline ($M=4.55, SD=1.23) (p=.001$).

\subsubsection{Non-HMD Observers}
The results portrayed in Figure \ref{SAM} demonstrate a significant difference in observers' perceived valence across the five conditions ($F_{2.172,41.268}=4.524, p=.015, \eta_p^2=0.192$). Post-hoc tests revealed that both VA ($M=4.85, SD=1.35) (p=.017$) and PO ($M=5.30, SD=1.30) (p=.003$) exhibited significantly higher valence scores compared to Baseline ($M=3.65, SD=1.53$).

For the arousal dimension, we found a significant effect of condition ($F_{2.586,49.136}=10.813, p<.001, \eta_p^2=0.363$). The post-hoc tests unveiled that the arousal score of Baseline ($M=3.10, SD=1.29$) was significantly lower than VA ($M=4.40, SD=1.50) (p=.036$), PO ($M=5.45, SD=1.15, p<.001$), PSW ($M=4.95, SD=1.36) (p=.003$), and PAR ($M=5.35, SD=1.66) (p=.001$).

\subsection{Interview Results}

\subsubsection{HMD users}
According to the average ranking score, the five conditions were ranked from best to worst performance as follows: PO (2.45), PAR (2.75), VA (3.05), PSW (3.20), and Baseline (3.55). Regarding their choices, 7 HMD users noted that PO \emph{``blends virtual and real well,''} and 6 found it \emph{``interesting.''} Additionally, 5 mentioned PO \emph{``provides helpful information for focus.''} For instance, P8 stated \emph{``only displays people without the surrounding environment, allowing better focus on the game and friend's reactions.''} For PAR, 4 HMD users praised its \emph{``realism and natural feel,''} while 3 appreciated the \emph{``broad field of view to see the real world.''} However, 5 found PAR \emph{``distracting due to excessive realism.''} Regarding PSW, 6 HMD users recognized its visibility but found it \emph{``unnatural in the virtual world.''} 5 participants likened it to \emph{``attending a remote meeting.''} Concerning VA, 3 HMD users found it \emph{``naturally integrated into the virtual world.''} However, 8 HMD users believed it \emph{``meaningless,''} with 5 stating that it is \emph{``difficult to connect this character with my friend.''} As for the Baseline, 10 HMD users considered it \emph{``boring''} and \emph{``lacking interaction,''} though 3 felt \emph{``more immersed in the game.''}

\subsubsection{Non-HMD Observers}

According to the average ranking score, the five conditions were ranked from best to worst performance as follows: PO (2.45), VA (2.70), PAR (2.80), PSW (3.05), and Baseline (4.00). In discussing their preferences, 6 observers favored PO, citing its \emph{``fun integration, like I'm entering the virtual world myself.''} 6 observers noted a \emph{``strong sense of participation''} in PO, while 3 mentioned that \emph{``showing only me on the screen doesn't interfere much with watching game content.''} For VA, 6 observers stated that appearing in a virtual avatar \emph{``looks natural''} and was \emph{``not awkward.''} However, 4 observers felt \emph{``a disconnection with the character.''}  In the case of PAR, 5 observers appreciated \emph{``a strong sense of participation,''} but 6 observers considered that \emph{``the realistic visuals are too distracting.''} Regarding PSW, 8 observers criticized its \emph{``unnatural integration''} and \emph{``disruption of game visuals.''} As for the Baseline, 13 observers labeled it as \emph{``boring''} and \emph{``lacking in involvement.''}

\begin{figure*}[htbp]
    \centering
    \includegraphics[width=\textwidth]{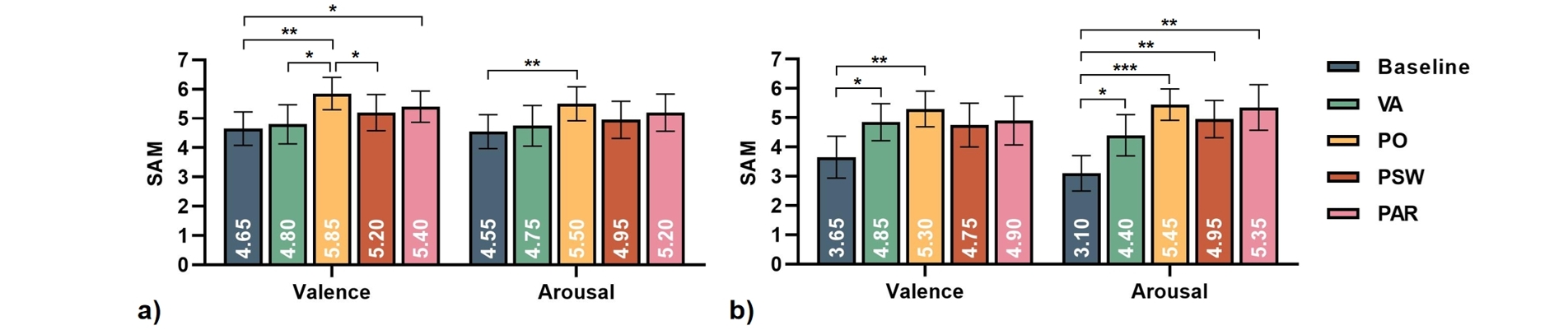}
    \caption{SAM ratings for (a) HMD users and (b) non-HMD observers in each condition. Error bars indicate 95\% confidence intervals.  *, **, and *** indicate statistical significance at the $p < 0.05, p < 0.01,$ and $p < 0.001$ levels, respectively.}
    \label{SAM}
\end{figure*}

\section{Discussion}
\subsection{Influence of Non-HMD Observer Representation on HMD User Performance and Experience during Co-located VR Sharing}
Contrary to \textbf{H1a}, we found that HMD users showed significant improvement in attention to the secondary task only in the PO condition compared to the Baseline. Contrary to \textbf{H1b}, users experienced a higher sense of co-presence in the Passthrough conditions compared to the Baseline, but not in the VA condition (see Figure \ref{Presence}). Our results also do not support \textbf{H1c}, as users exhibited more positive emotions only in the PO and PAR conditions compared to the Baseline  (see Figure \ref{SAM}). These results suggest that among these representations, PO is the most effective, consistent with HMD user interviews and ranking results. This effectiveness may be attributed to PO displaying the real view of the observer, enhancing focus and understanding during interaction \cite{cunningham2009dynamic}. Additionally, it minimizes distractions by excluding additional environmental elements.

Our results support \textbf{H2a}, we found that HMD users perceived more real-world information in the Passthrough conditions (see Figure \ref{Usability}). Our results also support \textbf{H2b}, as users felt that communication with observers was enhanced across all Passthrough conditions (see Figure \ref{Usability}). Contrary to \textbf{H2c}, the three Passthrough conditions led to a decrease in immersion compared to the Baseline, though there was no overall trend of decreased immersion compared to VA (see Figure \ref{Presence}). Despite reduced immersion, participants favored Passthrough conditions, suggesting minimal disruption to the gaming experience and highlighting its value in revealing the real world. Similar to findings on audio technologies \cite{o2022exploring}, HMD users prioritized awareness of observers over a sense of presence. The PSW condition, which also uses Passthrough, ranked lower due to its unnatural integration with the virtual world and its resemblance to remote meetings. Our findings affirm Passthrough's benefits for social interaction in shared spaces but suggest that presenting observers in a window format does not fully utilize its advantages.

In summary, HMD users favor representation methods with Passthrough, with PO being the favorite due to its effective integration with the virtual scene. It focuses on observers while preserving virtual content to minimize real-world distractions. Previous research \cite{gottsacker2021diegetic, von2019you} found that HMD users preferred virtual avatars for brief observer cues. However, our results indicated that for prolonged shared scenarios, perceiving real-life visuals of observers enhances the shared experience.



\subsection{Influence of Non-HMD Observer Representation on Observer Experience during Co-located VR Sharing}

Our results support \textbf{H3a}, we found that all four representations enhanced the observers' sense of presence in the virtual world (see Figure \ref{Presence}). Our results also support \textbf{H3b}, as observers showed more excited emotions under conditions with representations (see Figure \ref{SAM}). These findings highlight the positive impact of representations on a 2D screen for enhancing observers' shared experience. While prior research \cite{o2023re, gottsacker2021diegetic, o2022exploring, mcgill2015dose} has predominantly focused on the experiences of HMD users in shared spaces, our study marks an initial attempt to understand the perspectives and encounters of observers in these settings.

Our results support \textbf{H4a}, we found that Passthrough representations enabled observers to perceive more information (see Figure \ref{Usability}). Our results do not support \textbf{H4b}, as all three Passthrough representations caused significant disruption compared to the Baseline, but no overall trend was observed when compared to VA (see Figure \ref{Usability}). This disruption might stem from longer gaze durations on 2D screens compared to VR, as noted by \cite{loiseau2023exploring}. Therefore, displaying only the observers without the surrounding environment, as in PO, appears to be an optimal balance between realistic display and minimizing distractions, aligning with observers' reports in interviews.

In conclusion, representing observers on a 2D screen within the virtual environment enhances their watching experience and communication with HMD users. 
Since observers typically feel less engaged when merely watching \cite{sjoblom2017people}, representing them in the virtual environment can help. However, unlike HMD users who preferred realistic representations, observers had mixed preferences. Some liked realistic representations for stronger participation, while others preferred virtual avatars to avoid awkwardness. This discrepancy may relate to individual personality traits, highlighting the need for customizable design \cite{fenigstein1975public}.

\subsection{Design Guidelines}
We discovered that representing observers on both HMD and 2D screens positively addresses challenges for both HMD users, breaking their isolation, and helps observers enhance connection with the virtual world. However, their requirements for how to be represented differ. Therefore, our first recommendation is to provide distinct representations on HMD and 2D screens based on the specific needs of both parties. Following this, we provide design guidelines applicable to these two perspectives.

\subsubsection{Representing Non-HMD Observers on HMD}
HMD users prefer real visual representations of observers, allowing them to see real-time reactions and movements, enhancing co-presence and communication. We recommend displaying only the observer (PO) or the entire real environment (PAR) on HMDs. PO maintains more of the virtual environment and simulates observers entering the virtual space, while PAR merges virtual and real environments for a comprehensive real-world view. Displaying observers in a window format (PSW) is discouraged due to its unnatural integration, which reduces the sense of shared space. 


\subsubsection{Representing Non-HMD Observers on 2D Screens}
Unlike HMD users, observers watching on 2D screens preferred either a realistic representation or a virtual avatar, influenced by their personalities \cite{fenigstein1975public}. We recommend offering both options to accommodate different needs. Observers who preferred a realistic image felt it enhanced their participation and immersion. For this, we suggest using PO to show only the observer, as it best supports their sense of entering the virtual world. PAR is less necessary for observers already seeing the real environment, and PSW is not recommended due to its unnatural integration.

For those who preferred a virtual avatar, the benefits included reducing awkwardness and blending naturally with the virtual environment. We recommend using avatars consistent with the virtual world’s style and suggest two design enhancements: (1) offer customizable models for personalized avatars, and (2) use motion capture to align avatar movements with the observer's actions.


\subsection{Limitations and Future Work}
Our study provides valuable insights into enhancing shared experiences between HMD users and non-HMD observers. However, some limitations indicate future research directions. Firstly, our experiment involved a single observer, but real-life scenarios may include multiple observers. Future studies should explore how to adjust representation methods to manage visual interference from multiple observers.


We used the N-back task as a secondary task to test the attention of HMD users. Future research could explore real interactions between HMD users and observers. Additionally, to simulate their interaction, we had the observers deliver the cues, which to some extent affected the standardization of intervals.

We found that PO incurs an additional latency of approximately 50ms due to background removal processing. Although participants did not report experiencing latency issues during the experiment, this is considered a limitation of this condition that warrants attention in future research.

Lastly, participants noted that the VA method could be improved, citing issues with the virtual avatar's disconnection and lack of real-time animations. Future work should explore customizing avatar models and using motion capture for real-time animations to enhance the embodiment of avatars and observers.

\section{Conclusion}
Our study investigates the impact of non-HMD observer representation on both HMD users and non-HMD observers using HMDs and 2D screens. The findings show that HMD users prefer real-world visuals of observers, with PO being the favored method due to its minimal interference from the real world. Observers have varied preferences, favoring PO and VA for their realistic and virtual representations, respectively. We propose design guidelines to accommodate these differing perspectives, aiming to enhance co-located VR experiences. This research offers valuable insights for improving shared VR experiences for both HMD users and non-HMD observers.

\bibliographystyle{abbrv-doi}

\bibliography{template}
\end{document}